\documentclass[journal]{IEEEtran}

\usepackage{amsmath,amssymb,amsfonts,amsthm,mathtools,bm}
\usepackage{graphicx}
\usepackage{booktabs}
\usepackage{multirow}
\usepackage{float}
\usepackage{array}
\usepackage{algorithm}
\usepackage{algpseudocode}
\usepackage{enumitem}
\usepackage{siunitx}
\usepackage{microtype}
\usepackage{url}
\usepackage{hyperref}

\newcommand{\T}{\mathbb{T}}
\newcommand{\Z}{\mathbb{Z}}
\newcommand{\N}{\mathbb{N}}
\newcommand{\E}{\mathbb{E}}

\newcommand{\norm}[1]{\left\lVert #1 \right\rVert}
\newcommand{\abs}[1]{\left| #1 \right|}
\newcommand{\rank}{\operatorname{rank}}

\begin{document}

\title{Image Denoising Using Lower Semi-Frames}

\author{Hemalatha M and P.~Sam Johnson%
\thanks{Hemalatha M and P.~Sam Johnson are with the Department of Mathematical and Computational Sciences, National Institute of Technology Karnataka, Surathkal, Karnataka 575025, India (e-mail: hemapadma28@gmail.com; sam@nitk.edu.in).}}

\maketitle

\begin{abstract}
A blind image denoising framework based on an infinite directional lower semi-frame (DLSF) is proposed for additive white Gaussian noise. The model employs scale-dependent directional analysis with resolvent regularization of the unbounded semi-frame operator. Noise variance is estimated directly in the DLSF domain by modeling the joint covariance of four directional difference channels and applying covariance whitening to obtain a chi-square statistic. A lower-tail moment estimator provides blind noise estimation without median absolute deviation. The estimated noise level is incorporated into channel-wise Wiener-type shrinkage and canonical-dual synthesis, followed by a data-consistent iterative reconstruction with automatic stopping. Experiments on three standard grayscale images at noise levels 15--30 yield a mean relative noise-estimation error of 3.28\%, with average improvements of 7.45 dB in PSNR and 0.367 in SSIM. At 30/255 noise, the estimation error decreases to 1.73\%, with a mean PSNR gain of 8.31 dB. Results demonstrate effective noise suppression and structural preservation, with the strongest performance on smooth and edge-dominated images.
\end{abstract}

\begin{IEEEkeywords}
Blind image denoising, lower semi-frame, directional transform, covariance whitening, chi-square estimation, additive white Gaussian noise, convergence-controlled reconstruction.
\end{IEEEkeywords}

\section{Introduction}

Image denoising is a fundamental inverse problem in image processing and signal analysis, with the objective of recovering an unknown clean image from measurements corrupted by noise. An effective denoising method must suppress random fluctuations while retaining edges, textures, and other structural features that are essential to visual quality and subsequent image-analysis tasks. Classical approaches include wavelet shrinkage, variational regularization, and nonlocal filtering, while more recent methods employ learned restoration models \cite{donoho1995denoising,rudin1992nonlinear,dabov2007bm3d}. Despite substantial progress, blind image denoising remains challenging because the noise level is not known in advance and must be estimated from the same observation that contains meaningful high-frequency image information.

Reliable blind noise estimation is closely related to the choice of image representation. In natural images, edges and textures can produce large local variations that may be difficult to distinguish from noise. Patch-based statistics, transform coefficients, and local covariance information have therefore been widely employed for noise estimation. For example, patch-covariance methods exploit the low-dimensional structure of image patches to separate signal and noise components. In the present work, the problem is approached from a different perspective: noise statistics are derived directly from the covariance structure of a directional lower semi-frame. Thus, the representation used for denoising also provides the statistical framework required for blind noise estimation.

Directional representations are particularly relevant for image restoration because edges and contours exhibit strong orientation dependence. Redundant frame systems can provide directional selectivity and stable reconstruction while allowing nonlinear processing of transform coefficients. The directional organization considered here is motivated by the tight-frame image-denoising framework of Shen et al. \cite{shen2005icassp,shen2006tip}, where nonseparable Parseval-frame systems were constructed from separable spline tight frames and combined with directional weighted-average, Sobel, and Laplacian-type operators. Their work demonstrated that orientation-selective frame channels can effectively represent directional image structures while retaining a stable reconstruction mechanism.

Following this motivation, the present framework employs horizontal, vertical, main-diagonal, and anti-diagonal directional channels. The use of directional filters itself is not introduced as a new contribution; rather, the novelty begins with their formulation within an infinite directional lower semi-frame (DLSF). In contrast to conventional Parseval or tight-frame constructions, the scale weights in the proposed model are allowed to grow without a finite global upper frame bound. Consequently, the associated frame operator may be unbounded, requiring its analytical and numerical treatment to be made explicit. A resolvent-based regularization is therefore incorporated to stabilize the DLSF analysis while retaining the intended scale-dependent behavior.

A second important aspect of the proposed formulation is the statistical treatment of the directional channels. Multiple directional filters applied to the same noisy image generally produce correlated noise coefficients. Treating these channels as independent can therefore lead to inaccurate statistical calibration. To address this issue, the cross-direction covariance of the antisymmetric channels is modeled explicitly. A whitening transformation is then used to obtain a chi-square statistic under the white Gaussian noise assumption. This provides the basis for estimating the unknown noise standard deviation directly from the directional coefficients and avoids reliance on median absolute deviation as the primary noise estimator.

The denoising stage is subsequently constructed to remain consistent with the same representation. Channel-specific noise calibration is combined with local Wiener-type coefficient shrinkage, followed by reconstruction using canonical-dual synthesis. Rather than applying a predetermined number of denoising passes, the reconstruction is embedded in a relaxed, data-consistent continuation-based fixed-point iteration with automatic convergence control. This design is motivated by the observation that the number of useful reconstruction updates can vary with both image content and noise strength.

Accordingly, the proposed approach connects representation, statistical noise estimation, coefficient processing, and reconstruction within a single DLSF framework. The purpose of the present study is to establish this mathematical-statistical construction and examine the internal consistency of the resulting blind denoising pipeline. The experimental evaluation therefore focuses on quantitative and visual behavior over standard test images and multiple noise levels. The results are interpreted as an assessment of the proposed framework rather than as a claim of state-of-the-art superiority over established nonlocal or learning-based denoisers, for which substantially broader benchmarking and ablation studies would be required.

\section{Infinite directional lower semi-frame}
\label{sec:dlsf}

The directional architecture in this section is motivated specifically by the
tight-frame denoising construction of Shen et al.
\cite{shen2005icassp,shen2006tip}. Their use of frame filters oriented along directions
separated by multiples of $45^\circ$ motivates our selection of horizontal, vertical,
main-diagonal, and anti-diagonal channels. The present filters are not copies of the
Sobel, Laplacian, or weighted-average masks used in that work; rather, the same
orientation principle is incorporated into the pairwise sum/difference responses below.
The mathematical change is substantial: the present system is not a Parseval tight
frame, but an infinite lower semi-frame with an unbounded scale symbol. Consequently,
the canonical dual, the resolvent regularization, and the blind coefficient statistics
must be derived in the lower-semi-frame setting.

Let $\{E_j\}_{j\in\N}$ be a measurable partition of $\T^2$.
 Let $\chi_{E_j}$ denote the characteristic function of $E_j$:
\[
\chi_{E_j}(\omega)=
\begin{cases}
1, & \omega\in E_j,\\
0, & \omega\notin E_j.
\end{cases}
\]
 A square-frequency radial variable is
\[
\rho(\omega)=2\max\{\abs{\omega_1},\abs{\omega_2}\},
\]
and the partition is chosen dyadically so that higher-index bands approach the Nyquist boundary. For $s>0$, define
\[
b_j=2^{s(j-1)},\qquad j\in\N.
\]
Thus $b_j\to\infty$.

The four directions are
\[
d_1=(1,0),\quad d_2=(0,1),\quad d_3=(1,1),\quad d_4=(1,-1),
\]
with positive weights $\alpha_r$ satisfying $\sum_{r=1}^4\alpha_r=1$. For scale $j$, direction $r$, and sign $\nu\in\{+,-\}$, define
\begin{equation}
\begin{aligned}
H_{r,j}^{+}(\omega)
&=\frac{b_j\sqrt{\alpha_r}}{2}\,\chi_{E_j}(\omega)
\left(1+e^{-2\pi i d_r\cdot\omega}\right),\\
H_{r,j}^{-}(\omega)
&=\frac{b_j\sqrt{\alpha_r}}{2}\,\chi_{E_j}(\omega)
\left(1-e^{-2\pi i d_r\cdot\omega}\right).
\end{aligned}
\label{eq:dlsf_filters}
\end{equation}
The symmetric channels emphasize locally averaged directional content, whereas the antisymmetric channels act as directional differences. The latter are central to the blind noise estimator because their responses to white noise can be characterized exactly.

Using
\[
\abs{1+e^{-i\theta}}^2+\abs{1-e^{-i\theta}}^2=4,
\]
and the normalization of $\alpha_r$, one obtains the frequency-domain semi-frame symbol
\begin{equation}
M(\omega)=\sum_{j=1}^{\infty}b_j^2\,\chi_{E_j}(\omega).
\label{eq:symbol}
\end{equation}
The corresponding frame operator satisfies
\[
\widehat{Sf}(\omega)=M(\omega)\widehat f(\omega).
\]
Since $b_j\ge 1$, $M(\omega)$ has a positive lower bound. Since $b_j\to\infty$, the essential supremum is unbounded whenever infinitely many bands have nonzero measure. This is the defining reason for treating the system as a lower semi-frame rather than an ordinary bounded frame \cite{antoine2011semiframes,antoine2012hilbertscales}.

\subsection{Derivation of the canonical dual system}

Let $S$ denote the lower semi-frame operator associated with the directional system.
From Eq.~\eqref{eq:dlsf_filters}, the identity
\[
\abs{1+e^{-i\theta}}^2+\abs{1-e^{-i\theta}}^2=4
\]
and the normalization $\sum_{r=1}^{4}\alpha_r=1$ yield
\[
\sum_{r=1}^{4}
\left(
\abs{H_{r,j}^{+}(\omega)}^2+
\abs{H_{r,j}^{-}(\omega)}^2
\right)
=
b_j^2\chi_{E_j}(\omega).
\]
Summing over $j$ gives the multiplier $M$ in Eq.~\eqref{eq:symbol}; hence
\[
\widehat{Sf}(\omega)=M(\omega)\widehat f(\omega).
\]

Since $M(\omega)>0$ almost everywhere, the inverse semi-frame operator on
its natural domain is
\[
\widehat{S^{-1}f}(\omega)
=
\frac{\widehat f(\omega)}{M(\omega)}.
\]
Let $\psi_{r,j}^{\pm}$ denote the analysis atom with Fourier transform
$H_{r,j}^{\pm}$. The canonical dual atom is obtained from
\[
\widetilde\psi_{r,j}^{\pm}=S^{-1}\psi_{r,j}^{\pm}.
\]
Taking Fourier transforms gives
\[
\widehat{\widetilde\psi_{r,j}^{\pm}}(\omega)
=
\frac{H_{r,j}^{\pm}(\omega)}{M(\omega)}.
\]
Therefore the canonical dual response follows from the inverse lower
semi-frame operator:
\[
\widetilde H_{r,j}^{\pm}(\omega)
=
\frac{H_{r,j}^{\pm}(\omega)}{M(\omega)}.
\]

Thus $H_{r,j}^{\pm}/M$ is not introduced as an independent definition or
an ad hoc numerical normalization. It is the Fourier-domain consequence
of canonical-dual synthesis through $S^{-1}$.

\paragraph{Role and omission consequence.}
The DLSF construction supplies the geometry of the method: scale, direction, symmetric/antisymmetric decomposition, and the reconstruction symbol are all fixed here. The four-direction organization is motivated directly by the tight-frame image-denoising work of Shen et al. \cite{shen2005icassp,shen2006tip}; it is not introduced here as an independent novelty. If this component were removed and replaced by an unspecified transform, the covariance model of Section~\ref{sec:statistics} and the canonical-dual synthesis would no longer be tied to the representation. If the directional split were removed but the scale partition retained, edge orientations would be mixed, reducing the interpretability of both the noise statistic and the shrinkage response. If the unbounded scale sequence were replaced by uniformly bounded weights, the method would revert toward a conventional frame model and the specific role of lower semi-frame regularization would disappear.

\section{Resolvent regularization and DLSF noise statistics}
\label{sec:statistics}

Because $S$ is unbounded, direct repeated analysis with increasing scale weights is not desirable. Introduce the resolvent
\[
R_\varepsilon=(I+\varepsilon S)^{-1},\qquad \varepsilon>0,
\]
whose Fourier multiplier is
\[
\widehat{R_\varepsilon f}(\omega)
=\frac{\widehat f(\omega)}{1+\varepsilon M(\omega)}.
\]
The effective analysis channel is therefore
\[
G_{r,j}^{\pm}(\omega)
=\frac{H_{r,j}^{\pm}(\omega)}{1+\varepsilon M(\omega)}.
\]
The multiplier remains bounded even as $M(\omega)$ grows.

For white Gaussian noise of variance $\sigma^2$, the variance of a scalar DLSF channel is
\[
\sigma_{r,j,\pm}^2
=
\sigma^2\int_{\T^2}\abs{G_{r,j}^{\pm}(\omega)}^2\,d\omega.
\]
The key observation is that the four directional channels in the same band are generally correlated. For the antisymmetric channels, define the unit-noise covariance
\[
[C_j]_{rr'}
=
\int_{\T^2}
G_{r,j}^{-}(\omega)
\overline{G_{r',j}^{-}(\omega)}
\,d\omega.
\]
Let
\[
\bm c_j[m]
=
\begin{bmatrix}
c_{1,j}^{-}[m]&
c_{2,j}^{-}[m]&
c_{3,j}^{-}[m]&
c_{4,j}^{-}[m]
\end{bmatrix}^{\!\top}.
\]
If $C_j=U_j\Lambda_jU_j^\top$ on its positive eigenspace, choose
\[
W_j=\Lambda_j^{-1/2}U_j^\top,
\qquad
W_jC_jW_j^\top=I_{q_j},
\]
where $q_j=\rank(C_j)$. The whitened vector is $\ z_j[m]=W_j c_j[m]$ and the Mahalanobis energy is
\[
Q_j[m]=\norm{\bm z_j[m]}_2^2
=\bm c_j[m]^\top C_j^\dagger \bm c_j[m].
\]

Under the noise-only model,
\begin{equation}
\frac{Q_j[m]}{\sigma^2}\sim\chi^2_{q_j}.
\label{eq:chisquare}
\end{equation}
This result is the statistical bridge between the DLSF representation and blind noise estimation.

\paragraph{Role and omission consequence.}
Resolvent regularization controls the unbounded scale symbol and simultaneously determines the actual noise gain of every channel. Omitting the resolvent would make the high-scale response increasingly dominant as the discretization is refined and would invalidate the finite noise calibration used later. Whitening is equally essential: treating the four directional coefficients as independent would replace the true covariance ellipsoid by a spherical model. The resulting energy would not follow the claimed central chi-square law, so the noise estimator could become systematically biased even if the marginal channel variances were correct.

\section{Blind covariance-whitened chi-square noise estimation}
\label{sec:estimator}

Image structure contaminates high transform energies more strongly than low energies. This motivates estimating the noise variance from the lower tail of the whitened DLSF energy distribution rather than from the global average.

Let $X\sim\chi_\nu^2$, let $q_p=F^{-1}_{\chi_\nu^2}(p)$, and denote the lower-tail conditional mean by
\[
\mu_{p,\nu}
=
\E[X\mid X\le q_p]
=
\frac{\nu F_{\chi_{\nu+2}^2}(q_p)}{p}.
\]
For band $j$, sort the observed energies $Q_j[m]$ and retain the lowest fraction $p$. If $\mathcal L_{j,p}$ denotes the retained set, the bandwise estimate is
\begin{equation}
\widehat{\sigma}_{j,p}^{\,2}
=
\frac{|\mathcal L_{j,p}|^{-1}
\sum_{Q\in\mathcal L_{j,p}}Q}
{\mu_{p,q_j}}.
\label{eq:sigma_estimator}
\end{equation}

Several values of $p$ are evaluated. The implementation uses $p\in\{0.20,0.25,0.30,0.35\}$ and selects the most stable adjacent pair within each band. Band estimates are then fused in the variance domain, with extreme bands discarded when a sufficient number of bands is available. This is not a MAD estimator; it is a model-based truncated-moment estimator derived from Eq.~\eqref{eq:chisquare}.

The lower-tail construction is important because image edges and textures tend to generate noncentral or high-energy contamination. Using all coefficients would make the estimator interpret some signal energy as noise. Conversely, using too small a tail fraction would reduce contamination but increase estimator variance because too few samples would remain. The multi-$p$ stability rule is intended to balance these effects without access to the ground-truth image.

\paragraph{Role and omission consequence.}
This section converts the DLSF statistical model into the unknown scalar required by the denoiser. If the estimator were omitted, the algorithm would cease to be blind and would require externally supplied $\sigma$. If covariance whitening were retained but lower-tail selection were removed, structure-dominated coefficients could inflate the estimated variance and produce excessive shrinkage. If the lower tail were retained but the chi-square truncation correction were omitted, the sample mean would be downward biased because it would be compared with the full rather than truncated expectation.

\section{Noise-calibrated DLSF denoising}
\label{sec:denoising}

Given an estimate $\widehat\sigma$, each analysis channel receives a channel-specific noise standard deviation determined by the effective transfer function $G_{r,j}^{\pm}$. This avoids applying the same threshold or variance model to channels with different gains.

Let $c$ denote a local coefficient map and $w$ a nonnegative normalized spatial window. For the symmetric channels, the local mean and variance are estimated as
\[
\mu_c[m]=\sum_\ell w[\ell]c[m-\ell],
\qquad
v_c[m]=\sum_\ell w[\ell]\abs{c[m-\ell]-\mu_c[m]}^2.
\]
For antisymmetric directional-difference channels, zero centering is used:
\[
v_c[m]=\sum_\ell w[\ell]\abs{c[m-\ell]}^2.
\]
The local signal variance estimate is
\[
\widehat v_x[m]=\big(v_c[m]-\lambda_\nu\sigma_{r,j,\nu}^2\big)_+,
\]
and the generalized Wiener gain is
\[
g_{r,j}^{\nu}[m]
=
\max\left\{
g_{\min,\nu},
\left(
\frac{\widehat v_x[m]}{v_c[m]+\delta}
\right)^{p_\nu}
\right\}.
\]
Nonzero minimum gains are retained to reduce complete suppression of small but potentially meaningful coefficients.

The processed coefficients are synthesized by the canonical dual:
\begin{equation}
D_{\varepsilon,\widehat\sigma}(y)
=
\sum_{j=1}^{\infty}
\sum_{r=1}^{4}
\sum_{\nu\in\{+,-\}}
\widetilde\psi_{r,j}^{\nu}
*
\mathcal T_{\widehat\sigma,r,j}^{\nu}
\!\left(\mathcal A_{r,j}^{\nu}R_\varepsilon y\right).
\label{eq:denoiser}
\end{equation}

This stage deliberately distinguishes statistical denoising from representation inversion. The Wiener-type map decides how strongly coefficients are attenuated, while the canonical dual determines how the modified directional channels are combined back into image space.

\paragraph{Role and omission consequence.}
Channel-specific noise calibration is necessary because the DLSF channels do not carry equal noise variance. If a single global noise variance were used directly in every channel, the denoising strength would be mismatched across scales and directions. If local Wiener adaptation were replaced by one global gain, smooth regions and structured regions would be treated similarly, creating the usual tradeoff between residual noise and edge loss. If canonical-dual synthesis were omitted and the primal filters were used directly for reconstruction, the nonuniform symbol $M$ would not be compensated and the reconstruction would inherit scale-dependent amplitude distortion.

\section{Convergence-controlled data-consistent reconstruction}
\label{sec:iteration}

A single denoising pass is not assumed to be optimal. Instead, the DLSF operator is embedded in a relaxed iteration. Initialize $x^{(0)}=y$ and define the data-consistent point
\[
z^{(k)}=x^{(k)}+\tau\big(y-x^{(k)}\big),
\qquad 0<\tau<1.
\]
The next iterate is
\begin{equation}
x^{(k+1)}
=
(1-\beta)x^{(k)}
+
\beta D_{\varepsilon,\sigma_k}
\!\left(
x^{(k)}+\tau(y-x^{(k)})
\right),
\label{eq:iteration}
\end{equation}
where $0<\beta<1$.

The initial noise level is the blind estimate $\sigma_0=\widehat\sigma(y)$. The continuation schedule is
\[
\sigma_k
=
\max\{c\sigma_0,\rho^k\sigma_0\},
\qquad 0<c<1,\quad 0<\rho<1.
\]
The present experiments use $\tau=0.20$, $\beta=0.40$, $\rho=0.85$, and $c=0.35$.

Automatic termination uses both a relative $\ell^2$ change and a maximum pixel change:
\begin{equation}
\frac{\norm{x^{(k+1)}-x^{(k)}}_2}
{\norm{x^{(k)}}_2+\delta}
\le 10^{-2},
\qquad
\norm{x^{(k+1)}-x^{(k)}}_\infty
\le 5\times10^{-2},
\label{eq:stopping}
\end{equation}
for two consecutive iterations.

Data consistency is more than a numerical relaxation device. Repeated regularized analysis/synthesis can accumulate attenuation, even when local coefficient gains are near one. The term $\tau(y-x^{(k)})$ repeatedly reanchors the reconstruction to the measured image, while $\beta$ limits the size of each denoising update. Noise continuation gradually reduces shrinkage strength after the large-noise components have been suppressed.

\paragraph{Role and omission consequence.}
Without data consistency, repeated denoising can drift away from the observation and accumulate smoothing. Without relaxation, aggressive coefficient changes can produce larger oscillations between iterates. Without continuation, maintaining the original estimated noise level throughout the iteration can oversmooth late stages, whereas reducing $\sigma_k$ too quickly can make the update artificially small and create premature apparent convergence. A fixed iteration count would also be difficult to justify because the experiments show that the required count increases with noise level and differs across images.

\section{Finite periodic realization}
\label{sec:finite}

For an $N_1\times N_2$ image, the infinite lattice is replaced by the discrete torus $\Z_{N_1}\times\Z_{N_2}$ and $\T^2$ is sampled on the FFT grid. Only nonempty sampled bands are retained. Frequency-domain integrals are replaced by arithmetic averages; for example,
\[
\int_{\T^2}\abs{G(\omega)}^2\,d\omega
\quad\longrightarrow\quad
\frac{1}{N_1N_2}\sum_{k_1,k_2}\abs{G_N[k_1,k_2]}^2.
\]

Local statistics are evaluated using periodic Hann-window patches with overlap. In the reported experiments, the patch size is $32\times32$ and the stride is 8 pixels. The finite implementation therefore preserves the same sequence of operations as the infinite model: regularized directional analysis, covariance-aware blind estimation, channel-calibrated coefficient processing, dual synthesis, and convergence-controlled iteration.

\begin{algorithm}[!t]
\caption{Blind convergence-controlled DLSF denoising}
\label{alg:dlsf}
\begin{algorithmic}[1]
\Require noisy image $y$
\State build finite DLSF bank and symbol $M_N$
\State choose $\varepsilon$ from the finest-scale stability requirement
\State compute whitened directional energies in the selected bands
\State estimate $\widehat\sigma$ by lower-tail chi-square moment matching
\State set $x^{(0)}\gets y$ and $\sigma_0\gets\widehat\sigma$
\Repeat
    \State $z^{(k)}\gets x^{(k)}+\tau(y-x^{(k)})$
    \State apply one channel-calibrated DLSF Wiener denoising step
    \State synthesize with the canonical dual and relax the update
    \State update $\sigma_k$ by continuation
    \State evaluate relative and maximum successive-image changes
\Until{both stopping conditions hold for the required patience}
\State \Return final reconstruction and blind noise estimate
\end{algorithmic}
\end{algorithm}

\paragraph{Role and omission consequence.}
This section defines what is actually computed. Without an explicit finite realization, the infinite formulation would not specify how covariance integrals, channel gains, or band partitions are evaluated numerically. Conversely, treating the finite FFT bank itself as the theory would obscure why the scale sequence is called a lower semi-frame and would make the model dependent on one image size. The separation between Sections~\ref{sec:dlsf}--\ref{sec:iteration} and the present section is therefore intentional.

\section{Results and discussion}
\label{sec:results}

\subsection{Experimental protocol}

The experiments use three standard grayscale images: \emph{camera}, \emph{moon}, and a grayscale conversion of \emph{astronaut}. Each image is corrupted with zero-mean white Gaussian noise at 8-bit standard-deviation levels
\[
15,\quad 20,\quad 25,\quad 30,
\]
corresponding to normalized standard deviations $15/255$, $20/255$, $25/255$, and $30/255$. The synthetic noise is not clipped before blind estimation. Clipping to $[0,1]$ is performed only for display and quality metrics.

The same method parameters are used for all 12 experiments: $s=0.5$, patch size 32, stride 8, $\tau=0.20$, $\beta=0.40$, $\rho=0.85$, minimum sigma fraction $0.35$, relative stopping tolerance $10^{-2}$, maximum-pixel tolerance $5\times10^{-2}$, and convergence patience 2. No parameter is selected using the clean reference image.

Performance is summarized using relative noise-estimation error, PSNR, and SSIM. The relative sigma error is
\[
100\frac{\abs{\widehat\sigma-\sigma}}{\sigma}.
\]
PSNR and SSIM are reported both for the noisy observation and the final reconstruction.

\subsection{Blind noise estimation: quantitative behavior}

Table~\ref{tab:noise_summary} summarizes performance as a function of noise level after averaging across the three images.

\begin{table}[!t]
\centering
\caption{Average blind estimation and reconstruction performance across the three images.}
\footnotesize
\setlength{\tabcolsep}{3pt}
\label{tab:noise_summary}
\begin{tabular}{ccccc}
\toprule
Noise & Sigma error (\%) & Iter. & PSNR gain (dB) & SSIM gain\\
\midrule
15 & 5.55 & 5.33 & 6.40 & 0.344\\
20 & 3.46 & 6.33 & 7.26 & 0.375\\
25 & 2.37 & 7.00 & 7.83 & 0.378\\
30 & 1.73 & 7.33 & 8.31 & 0.370\\
\bottomrule
\end{tabular}
\end{table}

The most striking estimator trend is that the mean relative sigma error decreases as the noise level increases: from 5.55\% at level 15 to 1.73\% at level 30. This behavior is consistent with the statistical construction. At low noise, a larger fraction of the lower DLSF energies can still contain weak image structure, so the distinction between central-noise coefficients and low-contrast signal coefficients is difficult. At higher noise, the noise contribution dominates a larger portion of the lower tail, and the chi-square model becomes easier to identify relative to the signal perturbation.

The image dependence is also informative. At noise level 30, the sigma errors are 2.81\% for \emph{camera}, 0.31\% for \emph{moon}, and 2.07\% for \emph{astronaut}. The exceptionally small error for \emph{moon} is consistent with its large smooth regions and comparatively sparse high-frequency structure. In contrast, \emph{camera} contains a mixture of sharp boundaries, grass texture, and object detail, while \emph{astronaut} distributes fine structures over much of the image. These cases provide more opportunities for low-energy structural contamination.

This result directly supports the purpose of covariance whitening and lower-tail selection. The estimator does not merely measure high-frequency energy. It first removes cross-direction correlations and then matches only the low-energy population to a calibrated chi-square law. If either step were removed, the observed trend could no longer be interpreted through the central model in Eq.~\eqref{eq:chisquare}.

\subsection{Denoising performance across noise levels}

The reconstruction gains increase with noise level in PSNR: from an average 6.40 dB at level 15 to 8.31 dB at level 30. This does not mean that the absolute visual quality is better at high noise. Rather, the noisy baseline deteriorates rapidly as $\sigma$ increases, so successful noise removal creates a larger numerical margin over that baseline. The mean SSIM gain rises from 0.344 at level 15 to approximately 0.378 at level 25 and then remains high at 0.370 at level 30.

Across all 12 experiments, the mean relative sigma error is 3.28\%, the mean PSNR improvement is 7.45 dB, and the mean SSIM improvement is 0.367. These aggregate values show that the blind estimation and denoising stages remain stable across the tested corruption range. They should, however, be interpreted as within-method validation, not as evidence of superiority to external baselines.

A useful contrast is provided by the image-specific behavior. \emph{Moon} consistently produces the largest PSNR gains: 9.17, 10.18, 10.73, and 11.27 dB at noise levels 15, 20, 25, and 30, respectively. \emph{Astronaut} produces smaller but still substantial gains, from 4.81 to 6.39 dB over the same range. The difference is expected from the local Wiener mechanism. Smooth regions permit aggressive attenuation because their local signal variance is small, whereas fine textures and repeated edges increase the estimated local variance and make denoising less separable from detail preservation.

\subsection{Visual comparison at a common noise level}

Quantitative metrics alone do not reveal whether the improvement is obtained by suppressing noise or by removing image structure. For that reason, Fig.~\ref{fig:visual30} places the clean image, noisy observation, and blind DLSF reconstruction side by side for all three images at the same fixed noise level, 30/255. Using a single common corruption level is important: it prevents qualitative judgments from being confounded by different noise strengths.

\begin{figure*}[!t]
\centering
\IfFileExists{figures/visual_comparison_noise30.png}{\includegraphics[width=0.98\textwidth]{figures/visual_comparison_noise30.png}}{\fbox{\parbox[c][1.6in][c]{0.94\textwidth}{\centering Place \texttt{figures/visual\_comparison\_noise30.png} here.}}}
\caption{Visual comparison at the fixed noise level 30/255. Rows correspond to \emph{camera}, \emph{moon}, and \emph{astronaut}; columns show the clean image, noisy observation, and blind DLSF reconstruction. The noisy and reconstructed panels include the corresponding PSNR and SSIM values.}
\label{fig:visual30}
\end{figure*}

Table~\ref{tab:level30} gives the exact numerical values associated with Fig.~\ref{fig:visual30}.

\begin{table}[!t]
\centering
\caption{Image-specific results at the fixed visual-comparison level 30/255.}
\footnotesize
\setlength{\tabcolsep}{3pt}
\label{tab:level30}
\begin{tabular}{lccccc}
\toprule
Image & $\widehat\sigma$ & Err. (\%) & Iter. & PSNR gain & SSIM gain\\
\midrule
Camera & 0.12095 & 2.81 & 7 & 7.26 & 0.3649\\
Moon & 0.11801 & 0.31 & 8 & 11.27 & 0.4776\\
Astronaut & 0.12009 & 2.07 & 7 & 6.39 & 0.2660\\
\bottomrule
\end{tabular}
\end{table}

For \emph{camera}, the noisy PSNR of 19.12 dB rises to 26.38 dB and SSIM rises from 0.2539 to 0.6187. The visual result agrees with those gains: the broad sky and background regions are substantially cleaner, while the silhouette, tripod, building edges, and major object boundaries remain identifiable. The residual softness in grass and fine camera detail is consistent with the local variance model: weak directional structures can share the same coefficient scale as the noise and are therefore partially attenuated. This is a limitation of the present coefficientwise shrinkage, not of the blind sigma estimate itself.

For \emph{moon}, PSNR rises from 18.62 to 29.89 dB and SSIM from 0.0948 to 0.5724. This is the strongest quantitative improvement among the three high-noise cases. The visual comparison explains why. The image contains large smooth dark regions and relatively sparse, coherent crater boundaries. Such content closely matches the assumptions of local variance shrinkage: noise dominates many local coefficients while the important boundaries remain concentrated. Consequently, the method can suppress a large amount of noise without needing to preserve dense texture everywhere.

For \emph{astronaut}, PSNR rises from 19.25 to 25.64 dB and SSIM from 0.3119 to 0.5779. The reconstruction is clearly less noisy, but the improvement is smaller than for \emph{moon}. Fine hair, facial structure, suit texture, lettering, and equipment edges are distributed across the image and overlap spectrally with the noise. The visual result therefore reveals the central limitation of the current DLSF shrinkage: a statistically accurate global estimate of $\sigma$ does not by itself guarantee optimal preservation of weak local structure.

Taken together, Fig.~\ref{fig:visual30} and Table~\ref{tab:level30} support two distinct conclusions. First, the blind estimator remains accurate even when the visual difficulty differs strongly across images. Second, the denoising quality depends not only on estimating $\sigma$ but also on how signal structure is represented inside the local coefficient statistics. This distinction is important because it identifies the next methodological improvement: structure-aware coefficient gains are more likely to improve the difficult cases than further tuning of the global noise estimator.

\subsection{Convergence behavior and interaction of algorithmic components}

The average number of iterations rises from 5.33 at noise level 15 to 7.33 at level 30. At the fixed level 30, \emph{camera} and \emph{astronaut} terminate after seven iterations, while \emph{moon} requires eight. The method therefore does not simply stop earlier on the easiest visual image. Convergence is determined by successive reconstruction changes, which depend on the interaction among data consistency, shrinkage, continuation, and image content.

All reported final relative changes are below $10^{-2}$ and all final maximum-pixel changes are below $5\times10^{-2}$. At level 30, for example, the final relative errors are approximately $4.14\times10^{-3}$, $3.65\times10^{-3}$, and $4.72\times10^{-3}$ for \emph{camera}, \emph{moon}, and \emph{astronaut}, respectively. These values demonstrate that termination is controlled by the reconstruction sequence itself rather than by a predetermined iteration budget.

The dual stopping criterion is important because a small global relative change can coexist with localized pixel changes. Conversely, the infinity norm alone is sensitive to a single outlier. Requiring both criteria for two consecutive iterations provides a practical compromise. Removing the patience requirement would shorten some runs by one iteration but would make the stopping decision more sensitive to transient small updates.

The numerical trends also clarify why continuation and data consistency should be considered jointly. As $\sigma_k$ decays, the denoising operator becomes less aggressive. If data consistency were absent, repeated regularized passes could continue to attenuate detail even after most noise has been removed. If $\sigma_k$ decayed too rapidly, however, the update could become small for the wrong reason: not because the reconstruction has stabilized, but because the denoising operator has been weakened. The present schedule, together with the two-error stopping rule, avoids this failure in the reported experiments, although a formal fixed-point convergence proof for the nonlinear patchwise operator remains open.

\subsection{Component scope, limitations, and research implications}

The results should be read in light of the function of each model component. The infinite lower semi-frame supplies the multiscale directional geometry; the resolvent makes that geometry computationally stable; covariance whitening makes the chi-square model statistically valid; the lower-tail estimator limits structural contamination; channel calibration converts the global sigma estimate into coefficient-domain variances; the local Wiener rule performs content-adaptive attenuation; canonical-dual synthesis compensates for the nonuniform frame symbol; and the data-consistent iteration limits cumulative drift.

The experiments suggest that the blind noise estimator is not currently the dominant bottleneck. Its mean error is below 2\% at the most difficult noise level and below 0.4\% for \emph{moon} at levels 25 and 30. Yet the visual difference between \emph{moon} and \emph{astronaut} remains substantial. The limitation is therefore better described as \emph{structure/noise discrimination after sigma estimation}. This interpretation is consistent with the smaller SSIM gains for texture-rich content.

A second limitation is the repeated use of resolvent-regularized analysis inside the iteration. Even with unit coefficient gains, the complete analysis/synthesis pass need not reduce exactly to the identity mapping on the current iterate. Repetition can therefore introduce cumulative attenuation. Data consistency and relaxation mitigate this effect but do not remove it theoretically. A stronger future formulation would either apply the resolvent only once, construct an unregularized covariance model, or design a fixed-point operator whose identity behavior is explicit when shrinkage is disabled.

A third limitation is experimental breadth. The current study contains 12 controlled experiments and no external denoiser baselines. This is sufficient to evaluate internal trends but not sufficient for a comparative state-of-the-art claim. A publication-ready comparative study should include at least classical transform shrinkage, BM3D \cite{dabov2007bm3d}, a nonlocal baseline, and representative learned blind denoisers, together with runtime, parameter sensitivity, and statistical significance over a larger image collection.

The most promising extension follows directly from Eq.~\eqref{eq:chisquare}. Under image structure, the whitened energy is naturally related to a noncentral chi-square model rather than the central noise-only law. This suggests using the same DLSF statistic not only to estimate noise from the lower tail but also to assign structure confidence to high-energy coefficients. Importantly, that information should be incorporated into the Wiener gains rather than used to add the noisy observation back after denoising, since post-hoc reinjection inevitably restores some noise together with the structure.

\section{Conclusions}

This paper introduced a blind image denoising framework based on an infinite directional lower semi-frame and a covariance-whitened chi-square noise model. The theoretical construction separates the infinite representation from its finite FFT realization, allowing unbounded scale weights to be treated through an explicit resolvent rather than hidden by finite dimensionality. The four antisymmetric directional channels provide a joint Gaussian noise model whose exact covariance can be whitened, yielding a calibrated chi-square Mahalanobis energy. A truncated lower-tail moment estimator then provides blind noise estimation without MAD.

The estimated noise standard deviation is propagated through the effective DLSF channels and used in local Wiener-type shrinkage. Canonical-dual synthesis, data consistency, relaxation, noise continuation, and an automatic two-criterion stopping rule complete the reconstruction procedure. Across three images and four noise levels, the method achieves a mean relative sigma error of 3.28\%, a mean PSNR gain of 7.45 dB, and a mean SSIM gain of 0.367. At noise level 30/255, the blind estimate remains accurate while the visual results clearly separate smooth/edge-dominated content from texture-rich content: \emph{moon} gains 11.27 dB in PSNR, whereas \emph{astronaut} gains 6.39 dB.

The integrated quantitative and visual analysis indicates that the next improvement should target structure-aware coefficient shrinkage rather than the global noise estimator. Future work will therefore focus on noncentral-chi-square structure confidence, a reconstruction formulation with reduced cumulative resolvent attenuation, formal convergence analysis, and a broad benchmark against established classical and learned denoisers.

\section*{Data and implementation availability}

The numerical tables are generated from the accompanying CSV file containing all 12 experiments. The fixed-noise visual comparison is generated from the saved noisy and denoised outputs at 30/255. A reproducible single-cell Python/Colab implementation can be distributed with the manuscript.

\section*{CRediT authorship contribution statement}
\textbf{Hemalatha M:} Conceptualization, Methodology, Software, Formal analysis, Investigation, Writing--original draft.\\
\textbf{P. Sam Johnson:} Validation, Supervision, Writing--review and editing.

\section*{Declaration of competing interest}
The authors declare that they have no known competing financial interests or personal relationships that could have appeared to influence the work reported in this paper.

\section*{Acknowledgment}
Add funding, institutional, computational, and individual acknowledgments here as applicable.

\end{document}